\pdfoutput=1
\documentclass[11pt]{article}

\usepackage[final]{acl}
\usepackage{comment}
\usepackage{times}
\usepackage{latexsym}
\usepackage{adjustbox}
\usepackage{multirow}
\usepackage{amsmath}
\usepackage{graphicx}
\usepackage{subcaption}
\usepackage{amsmath}
\usepackage{amssymb}
\usepackage[T1]{fontenc}
\usepackage[utf8]{inputenc}

\usepackage{microtype}

\usepackage{inconsolata}

\title{ Technical report: Impact of Duration Prediction on Speaker-specific TTS for Indian Languages}

\author{
  Isha Pandey\textsuperscript{1} \quad
  Pranav Gaikwad\textsuperscript{2} \quad
  Amruta Parulekar\textsuperscript{1} \quad
  Ganesh Ramakrishnan\textsuperscript{1} \\
  \textsuperscript{1} Indian Institute of Technology Bombay, India \\
  \textsuperscript{2} BharatGen, India \\
}
\begin{document}
\maketitle
\begin{abstract}
High-quality speech generation for low-resource languages, such as many Indian languages, remains a significant challenge due to limited data and diverse linguistic structures. Duration prediction is a critical component in many speech generation pipelines, playing a key role in modeling prosody and speech rhythm. While some recent generative approaches choose to omit explicit duration modeling, often at the cost of longer training times. We retain and explore this module to better understand its impact in the linguistically rich and data-scarce landscape of India.

We train a non-autoregressive Continuous Normalizing Flow (CNF) \cite{cfm} based speech model using publicly available Indian language data and evaluate multiple duration prediction strategies for zero-shot, speaker-specific generation. Our comparative analysis on speech-infilling tasks reveals nuanced trade-offs: infilling based predictors improve intelligibility in some languages, while speaker-prompted predictors better preserve speaker characteristics in others. These findings inform the design and selection of duration strategies tailored to specific languages and tasks, underscoring the continued value of interpretable components like duration prediction in adapting advanced generative architectures to low-resource, multilingual settings.

% \begin{itemize}
%     \item Using cross-attention for frame-to-text alignment bypasses duration prediction, leveraging monotonicity inherent in speech-text alignment. \textcolor{purple} ByT5 hindi pre-trained encoder to get text features. - Only on Hindi and English. 
%     \item Baselines
%     \begin{itemize}
%         \item F5TTS - Convergence takes too long and prosody/accent is not good
%         \item Voicebox -  duration Predictor
%         \item Xtts2 - for style transfer
%     \end{itemize} 
%     \item What capabilities we should show?
%     \begin{itemize}
%         \item style transfer - Continuation/Cross Sentence
%         \item \textcolor{red}{Content editing} - maybe add in demo 
%         \item \textcolor{red}{Diverse Sampling - not explored}
%     \end{itemize}
    
% \end{itemize}
  
\end{abstract}

\section{Introduction}
% About advancements in tts and the need for zero shot speaker specific tts
%existing for english and some other languages but
%about indian languages and speakers and their diversity and no such model existing for zero shot speaker conditioning.. ones that exist though multispeaker 
 
%our contributions duration predictor, opens up tasks like speech infilling 

Large-scale generative models have made remarkable progress across domains, but achieving high-quality and context-sensitive speech generation for diverse linguistic landscapes remains a challenge, particularly for languages with rich morphology, significant dialectal variation, and phonetic structures that diverge considerably from high-resource languages like English. Indian languages exemplify this complexity. Directly applying architectures developed for high-resource settings to Indian languages is often ineffective due to their intricate prosody, agglutinative or inflectional morphology, and the general lack of standardized orthographies across dialects. Moreover, state-of-the-art speech generation systems are typically trained on massive datasets, often exceeding 60,000 hours, whereas our work investigates these challenges using a much smaller dataset of approximately 2,000 hours or less, reflecting the true low-resource conditions common to many Indian languages.

In this work, we conduct a focused investigation into the role of duration prediction within the speech generation pipeline, aiming to better understand its influence on prosody and speaker-related features in Indian languages. As described in Section \ref{sec:probelm_formualtion}, we employ a Continuous Normalizing Flow (CNF)–based audio model inspired by the Voicebox architecture\cite{cfm,le2023voiceboxtextguidedmultilingualuniversal} and implement two distinct duration prediction strategies for comparison. Rather than optimizing solely for end-task performance, our goal is to explore how architectural choices particularly in duration modeling interact with the linguistic diversity and low-resource realities of Indian speech, offering insights that may guide future model design and adaptation.

Our study centers on adapting the duration prediction mechanism within the Voicebox-inspired framework. In typical systems such as Voicebox (see Section \ref{sec:infill_duration_prediction}), durations are predicted based on text and explicit timing targets obtained from a pre-trained forced alignment model. However, this approach can be unreliable in low-resource contexts like Indian languages, which exhibit rich prosodic variation—including rhythm, stress, and intonation. Forced alignments may fail to capture such variation accurately. To address this, we adopt a prompt-based duration prediction strategy inspired by recent work in prosody modeling, such as PFlow\cite{kim2023pflow}. As discussed in Section \ref{sec:duration_prediction}, this approach replaces external alignment inputs with a three-second speaker prompt, which—when processed using cross-attention—enables the model to extract speaker-specific prosodic cues directly from reference audio. This design allows for more contextually appropriate and natural duration predictions and better captures the expressive prosody characteristic of Indian languages. Our adaptation aims to evaluate the feasibility and impact of prompt-based duration modeling in a multilingual, low-resource setting where alignment-dependent techniques often struggle.

For the audio generation component, detailed in Section \ref{sec:audio_model}, we adopt the non-autoregressive CNF-based modeling approach used in Voicebox. This enables the transformation of a simple latent distribution into the complex distribution of missing speech segments, conditioned on both text and surrounding audio context, i.e., modeling \( p(\text{missing data} \mid \text{context}) \). Training is performed using the flow-matching objective, which supports efficient and scalable optimization of CNFs via vector field regression. While the audio model architecture closely follows Voicebox, we modify the training strategy to improve learning robustness given the limited and diverse nature of the Indian language data.

We evaluate the two proposed duration prediction strategies through detailed empirical analysis, using datasets described in Section \ref{dataset}. As detailed in Section \ref{sec:results}, our experiments focus on speech infilling tasks, including both continuous sentence completion and cross-sentence completion, conducted across multiple Indian languages. To assess performance, we use Word Error Rate (WER) and the SIMO metric to evaluate intelligibility and similarity to ground-truth speaker characteristics, respectively. Additionally, we conduct human evaluations to complement these objective metrics. The results reveal language-dependent trade-offs: in some cases, prompt-based predictors improve intelligibility, while in others, they better preserve speaker-specific prosodic traits.These observations highlight how prosody, language structure, and modeling decisions collectively influence speech generation performance across diverse Indian languages.

% The contribution of this work can be summarized as follows:

% \begin{itemize}
%     \item 
%     \item 
%     \item
% \end{itemize}

\section{Related work}
\textbf{Foundations in Large-Scale Generative Speech Models:} There have been several advancements in zero-shot text-to-speech (TTS) in the recent years. Our work is primarily based on Voicebox \cite{le2023voiceboxtextguidedmultilingualuniversal}, which is a text-guided multilingual speech generation model that uses non-autoregressive flow-matching for speech infilling, given audio context and text. It can perform a variety of tasks inlcuding noise removal, style transfer and zero-shot TTS conversion. Flow matching \cite{lipman2023flowmatchinggenerativemodeling} is a framework for training continuous normalization flows by regressing a vector field that translates a source distribution to a target data distribution along predetermined probability paths. This simulation-free approach allows for faster training, sampling, and better generalization than diffusion paths. There exist several other architectures for text-to-speech synthesis. NaturalSpeech \cite{tan2022naturalspeechendtoendtextspeech} uses a variational autoencoder for end-to-end text-to-speech modeling with several strategies like phoneme pretraining, differentiable duration modeling, bidirectional prior-posterior modeling, and a memory mechanism to improve the speech quality and minimize artifacts. Vall-E \cite{wang2023neuralcodeclanguagemodels} uses a neural codec audio model with discrete audio codec tokens to generate high quality personalized zero-shot speech by framing TTS as a conditional language modeling task. Tacotron 2 \cite{shen2018naturalttssynthesisconditioning} uses a recurrent sequence-to-sequence feature prediction network to map character embeddings to mel spectrograms and generate high-quality speech from input text. WaveNet \cite{oord2016wavenetgenerativemodelraw} is a fully probabilistic and autoregressive deep generative model for natural speech synthesis, for which each audio sample's predicted distribution is conditioned on all previous audios.

\noindent\textbf{Speech Generation for Low-Resource Languages, Particularly Indian Languages:} Despite the fast developments on zero-shot text-to-speech synthesis in high resource languages, the literature for low-resource, particularly Indian languages, is limited. \cite{Panda2020ASO} highlights that the technological advancements for TTS reflect for less than 60 \% of the official languages of India while research for the unofficial languages is yet to begin. The paper also highlights the challenges faced in designing Indian Language TTS systems due to the linguistic variations in these languages. IndicTTS \cite{kumar2023buildingtexttospeechsystemsbillion} was the first study to train and evaluate TTS systems based on Transformers \cite{vaswani2023attentionneed} and HifiGAN \cite{kong2020hifigangenerativeadversarialnetworks} on Indian languages. However, it does not include multi-speaker and speaker-specific TTS. Since then, several new multi-speaker datasets such as IndicVoices-R \cite{sankar2024IndicVoicesrunlockingmassivemultilingual} have been released, and it is hence crucial to use these datasets and develop speaker-specific TTS systems for Indian languages.

\noindent \textbf{Innovations in Duration Prediction and Prosody Modeling:} Duration prediction is a popular method used to convert text to natural sounding speech. FastSpeech \cite{ren2019fastspeechfastrobustcontrollable} extracts attention alignments from an encoder-decoder based teacher model for phoneme duration prediction. These durations are then used by a length regulator to expand the source phoneme duration to match the target mel spectrogram. VITS \cite{kim2021conditionalvariationalautoencoderadversarial} proposes a stochastic duration predictor to synthesize speech with diverse rhythms from input text, emphasizing that the same text can be naturally spoken in different ways. Non-attentive Tacotron \cite{shen2021nonattentivetacotronrobustcontrollable} replaces the attention-based alignment mechanism in Tacotron 2 \cite{shen2018naturalttssynthesisconditioning} with an explicit duration predictor which allows for utterance-wide and per-phoneme duration control at inference time. Apart from direct duration prediction, other strategies have also emerged for prosody and style transfer. Style tokens \cite{wang2018styletokensunsupervisedstyle} uses embeddings that are jointly trained with Tacotron \cite{shen2018naturalttssynthesisconditioning} and are able to learn a wide range of expressiveness. There have also been RAD-TTS \cite{shih2021radtts} uses normalizing flows, robust alignment learning and models speech rhythm as a separate generative distribution to enable improved prosody transfer. For our work, we derived inspiration from these duration modeling techniques to develop a highly effective duration predictor that can work without relying on external forced alignments.

\noindent\textbf{Non-Autoregressive Audio Modeling:} Our audio model was based on Voicebox \cite{le2023voiceboxtextguidedmultilingualuniversal}, but there are several other TTS models that use non-autoregressive (NAR) modeling, such as FastSpeech \cite{ren2019fastspeechfastrobustcontrollable} and FastSpeech 2 \cite{ren2022fastspeech2fasthighquality}. Hifi-GAN \cite{kong2020hifigangenerativeadversarialnetworks} is a state-of-the-art NAR vocoder that models the periodic patterns in audio. More recently, flow-based NAR TTS models like Glow-TTS \cite{kim2020glowttsgenerativeflowtexttospeech} have emerged, which does not require any external alignments. Finally, EfficientTTS \cite{miao2020efficientttsefficienthighqualitytexttospeech} focuses on efficient NAR speech generation by using a new monotonic alignment modeling approach.

\section{Methodology}
\subsection{Background}
Flow Matching with Optimal Transport
Continuous Normalizing Flows (CNFs) provide a powerful framework for learning complex data distributions by transforming a simple prior distribution $p_0$ to a target data distribution $p_1$. This transformation is achieved through a time-dependent vector field $v_t : [0, 1] \times \mathbb{R}^d \to \mathbb{R}^d$, which constructs a flow $\phi_t$ governed by the ordinary differential equation:

$\frac{d}{dt}\phi_t(x) = v_t(\phi_t(x)) ; \phi_0(x) = x$ 
For a given flow $\phi_t$, we can derive the probability path pt(x) using the change of variables formula: 
$p_t(x) = p_0(\phi^{-1}_t(x)) \left|\det\left(\frac{\partial\phi^{-1}_t}{\partial x}(x)\right)\right|$
To train the neural network parameters $\theta$ that define our vector field $v_t(x; \theta)$, we employ the Flow Matching objective: 

% \[ \mathcal{L}_{FM}(\theta) = \mathbb{E}_{t,p_t(x)} \left\| u_t(x) - v_t(x; \theta) \right\|^2 \]

\[
\mathcal{L}_{FM}(\theta) = \mathbb{E}_{t,p_t(x)} \left\lVert u_t(x) - v_t(x; \theta) \right\rVert^2
\]

However, directly computing this objective is challenging as we lack prior knowledge of pt or $v_t$. To address this, we utilize the Conditional Flow Matching (CFM) objective, which provides a practical training approach: 

% \[\mathcal{L}_{CFM}(\theta) = \mathbb{E}_{t,q(x_1),p_t(x|x_1)} \left\|u_t(x|x_1) - v_t(x; \theta) \right\|^2\]

\[
\mathcal{L}_{CFM}(\theta) = \mathbb{E}_{t,q(x_1),p_t(x|x_1)} \left\lVert u_t(x|x_1) - v_t(x; \theta) \right\rVert^2
\]

For our implementation, we specifically adopt the optimal transport (OT) path, which defines the conditional probability and vector field as:
$p_t(x|x_1) = \mathcal{N}(x|tx_1,(1-(1-\sigma_{min})t)^2I)$  and 
$u_t(x|x_1) = \frac{x_1-(1-\sigma_{min})x}{1-(1-\sigma_{min})t}$

The OT path is particularly advantageous as it ensures points move with constant speed and direction, leading to more stable training and efficient inference. This choice simplifies the learning process while maintaining the model's expressive power.

\subsection{Problem Formulation and Model Training} 
\label{sec:probelm_formualtion}

Our work adapts the approach from Voicebox \cite{le2023voiceboxtextguidedmultilingualuniversal} to Indian languages. Given speech-transcript pairs $(x, y)$, we train a model for text-guided speech generation through in-context learning. The model learns speech infilling - predicting missing speech segments given surrounding audio and text transcript. Using a binary mask $m$, we split the audio into missing ($x_{mis}$) and context ($x_{ctx}$) portions, training the model to learn $p(x_{mis} | y, x_{ctx})$.
Following voicebox \cite{le2023voiceboxtextguidedmultilingualuniversal}, we use an audio model and a duration predictor for fine-grained alignment control

The system works with audio frames $x = (x_1,...,x_N)$, character sequence $y = (y_1,...,y_M)$, and per-character durations $l = (l_1,...,l_M)$. The frame-level transcript $z$ is derived by repeating each character $y_j$ according to its duration $l_j$. For any input pair $(x, y)$, we obtain $l$ and $z$ through forced alignment using a speech recognition model. The final prediction combines the audio model $q(x_{mis} | z, x_{ctx})$ and duration model $q(l_{mis} | y, l_{ctx})$, where $l_{mis}$ and $l_{ctx}$ represent masked and unmasked portions of the duration sequence.
Additionally, inspired by PFlow [citation], we explored an alternative duration predictor that takes a variable-length unmasked audio portion (minimum 2 seconds) to predict the duration distribution for each 
character in the text. This approach learns $q(l_{mis} | y, x_{ctx})$ directly, eliminating the need for explicit duration context $l_{ctx}$.

\subsection{Audio Model}
\label{sec:audio_model}

Following voicebox \cite{le2023voiceboxtextguidedmultilingualuniversal}, we implement a Continuous Normalizing Flow (CNF) model that learns the distribution of missing audio frames given the context. The audio is represented as an 80-dimensional log Mel spectrogram ($x_i \in \mathbb{R}^{80}$).
For a given input of context spectrogram $x_{ctx} \in \mathbb{R}^{N \times F}$, flow state $x_t \in \mathbb{R}^{N \times F}$, character sequence $z \in [K]^N$ (where $K$ is the number of character classes), and time step $t \in [0, 1]$, we employ a Transformer to parameterize the vector field $v_t$. The model is trained using a masked version of the flow matching objective:

% \begin{align}
%     L_{audio-CFM}(\theta) &= \mathbb{E}_{t,m,q(x,z),p_0(x_0)} \left\lVert u_t(x_t|x) \nonumber \\
%                      - v_t(x_t,x_{ctx},z; \theta)\right\rVert^2
% \end{align}

\begin{align}
    L_{audio-CFM}(\theta)
    &= \mathbb{E}_{t,m,q(x,z),p_0(x_0)} \left\lVert u_t(x_t|x) \right. \nonumber \\
    &\quad - \left. v_t(x_t,x_{ctx},z; \theta) \right\rVert^2
\end{align}

Where the masked context $x_{\text{ctx}}$ is defined as:
\[
x_i^{\text{ctx}} = 
\begin{cases}
0 & \text{if } m_i = 1, \\
x_i & \text{if } m_i = 0,
\end{cases}
\]
,
And the binary mask $m_i$ identifies the frames to be predicted (masked portion) versus the available context frames (unmasked portion).
The input to the Transformer is then constructed by concatenating the three sequences $(x_t, x_{\text{ctx}}, z_{\text{emb}})$ along the feature dimension:
\[
H_c = \text{Proj}([x_t; x_{\text{ctx}}; z_{\text{emb}}]) \in \mathbb{R}^{N \times D},
\]
where $D$ is the Transformer embedding dimension.

To condition on the flow step $t$, a sinusoidal positional encoding maps $t \in [0,1]$ to a vector $h_t \in \mathbb{R}^D$. The final input to the Transformer is $H_c$ concatenated with $h_t$:
\[
\widetilde{H}_c \in \mathbb{R}^{(N+1) \times D}.
\]

The transformer outputs $v_t(x_t, x_{\text{ctx}}, z; \theta) \in \mathbb{R}^{N \times F}$, which corresponds to the original sequence length $N$.

Unlike Voicebox training, our loss function interpolates between the masked and unmasked portions, applying a weighted sum. Specifically, we assign a weight of 0.9 to the loss computed on the masked frames and a weight of 0.1 to the loss on the unmasked (context) frames. This strategy allows the model to prioritize learning the content of the segments that require prediction while simultaneously being regularized to maintain the integrity and quality of the provided audio context.

\subsection{Duration Prediction}
\label{sec:duration_prediction}

The Voicebox model \cite{le2023voiceboxtextguidedmultilingualuniversal} requires accurate character durations for high-quality speech generation. For Indian languages, predicting these durations presents unique challenges due to the rich diversity in regional pronunciations and speaking styles. Through extensive experimentation, we explored different approaches to duration prediction, analyzing their impact on both intelligibility and speaker similarity in the generated speech.

\begin{table*}[h]
\small{
\centering
\begin{tabular}{|l|c|c|c|}
\hline
\textbf{Dataset} & \textbf{\# Speakers} & \textbf{\# Lang.} & \textbf{Lang. Scope} \\
\hline
IndicVoices \cite{javed2024IndicVoicesbuildinginclusivemultilingual} & 22563 & 22 & Indic only \\
IndicSuperb \cite{javed2022IndicSuperbspeechprocessinguniversal} & 1218 & 12 & Indic only \\
IndicTTS \cite{kumar2023buildingtexttospeechsystemsbillion} & 44 & 22 & Indic only \\
Spring \cite{r2023springinxmultilingualindianlanguage}& 7609 & 10 & Indic only \\
FLEURS \cite{conneau2022fleursfewshotlearningevaluation} & Not specified & 102 & International \\
Mucs \cite{Diwan-2021} & Not specified & 6 & Indic only \\
Shrutilipi \cite{DBLP:conf/icassp/BhogaleRJDKKK23} & Not specified & 12 & Indic only \\
SYSPIN \cite{abhayjeet2025syspins1} & 18 & 9 & Indic only \\
Common Voice \cite{ardila2020commonvoicemassivelymultilingualspeech} & Not specified & 60 & International \\
Vaani \cite{vaani2025} & 112394 & 54 & Indic only \\
Dhwani \cite{sanket2025IndicST} & Not specified & 40 & Indic only \\
Bhashini (Multiple datasets) & Not specified & 19 & Indic only\\
OpenSLR (Multiple datasets) & Not specified & 25 & International\\
\hline
\end{tabular}
\caption{Number of Speakers, Languages, and Language Scope in Various Speech Datasets}
\vspace{-10pt}
\label{tab:speech_datasets}
}
\end{table*}

\subsubsection{Voicebox style infilling Duration Predictor}
\label{sec:infill_duration_prediction}

The duration predictor which was used in Voicebox \cite{le2023voiceboxtextguidedmultilingualuniversal} was used to predict durations from the text and feed them to the audio model.
Inspired by the audio model, it models $q(l | y, l_{\text{ctx}})$ using a conditional vector field. $l$ represents duration, $l_{\text{ctx}}$ is the context duration, $y$ is the phonetic transcript.
It swaps $(x, x_{\text{ctx}}, z)$ from the audio model with $(l,l_{\text{ctx}},y)$. Training is done using a masked version of the Conditional Flow Matching (CFM) loss.
\par
While we use the regression-based duration modeling approach promoted by them where they regress the masked duration $l_\text{miss}$ given $l_\text{ctx}$ and $y$. In our implementation, learnable embeddings of the text characters and corresponding durations are concatenated along the frame dimension. This concatenated representation is then projected down using a linear projection layer. To capture local temporal patterns, we introduce a convolutional network before passing the features to the transformer. The Transformer architecture is similar to the audio model with fewer parameters. Unlike Voicebox \cite{le2023voiceboxtextguidedmultilingualuniversal},  in our implementation we train duration predictor with MSE loss over log-scaled durations over masked phones.

\subsubsection{Speaker-Prompted Duration Predictor}
Accurately modeling phoneme durations is critical for synthesizing natural-sounding speech, yet presents a significant challenge in the context of Indian languages. This difficulty is exacerbated by the high variety of dialects and pronunciations across different regions of the country, which introduces considerable variability in phoneme durations within our training data. To address this high variance and improve the robustness and generalization capability of duration prediction under data constraints, we developed an enhanced duration predictor. This design was inspired by approaches that leverage speech prompts for conditioning, such as the method explored in PFlow \cite{kim2023pflow}.

Unlike conventional approaches that condition duration prediction directly on text and rely on forced aligned durations as contextual input, which can be unreliable in low-resource settings and often lead to unnatural prosody, we treat these alignments as weak supervision during training. Rather than conditioning the model on hard-aligned durations, we used a 3-second mel spectrogram segment $x_p$, randomly sampled from the context mel spectrogram $x_{\text{ctx}}$, together with the text sequence $c$ as input. This allows the model to implicitly learn prosodic patterns and speaker-specific durations from real mel prompt, guided by weak alignment signals, without depending on them explicitly at inference time.

The core of our model is a transformer-based encoder $f_{\text{enc}}$ that produces a speaker-conditioned text representation $h_c = f_{\text{enc}}(x_p, c)$. The encoder embeds the text sequence $c$ using learnable embeddings and projects the mel-spectrogram $x_p$ via a linear layer to match the text embedding dimensionality. Cross-attention is then applied from the text tokens $c$ to the speech frames in $x_p$, enabling the model to fuse prosodic cues into the text representation.

The resulting representation $h_c$ is passed through two feed-forward layers with non-linear activation to predict log-duration values for each token in the input sequence of length $N$. The model is trained using a mean squared error (MSE) loss between the predicted and ground-truth log-scaled durations:
\[
\mathcal{L}_{\text{dur}} = \frac{1}{N} \sum_{i=1}^N \left( \log \hat{d}_i - \log d_i \right)^2,
\]
where $\hat{d}_i$ and $d_i$ are the predicted and target durations for the $i$-th token, respectively.

The audios generated by the Voicebox using these predicted durations were considerably better in both intelligibility and speaker similarity compared to using durations from an external forced alignment module.

\section{Experimental Setup}
\label{sec:experiments}

\subsection{Datasets}
\label{dataset}
Several transcribed Indian language datasets were used for training the model to achieve a balanced representation of urban and rural speakers and recording devices. This created a diversity in the vocabulary, content, and recording channels, including a mix of read speech, voice commands, extempore discussions, and both wide and narrow-band recordings. Table \ref{tab:speech_datasets} depicts the details of the different publically available datasets used for training the model. 

% We have used publically available open-source datasets like Bhashini, IndicSuperb, IndicVoices, SYSPIN, Fluers, Shrutilipi and OpenSLR. %\textcolor{green}{No point in adding descriptions, we should just cite in table and add info like number of speakers/no. of hours etc.}

\subsection{Data Preprocessing}
\subsubsection{Text Normalization:}
The transcripts, being from multiple datasets, had inconsistencies and were normalized to remove punctuation, spaces and other characters. Sentences containing words in the latin script were also removed from the Fluers and Spring datasets to create homogenized monolingual datasets for each target language.
\subsubsection{Audio Filtering:}
Some audios were unintelligible, noisy, or had transcription errors. To remove such audios, all audios were transcribed using the IndicWhisper \cite{bhogale2023vistaardiversebenchmarkstraining} model and Word Error Rate (WER) of these transcripts with the ground-truth transcripts was calculated. All audios with WER more than 0.2 were discarded as low-intelligibility audios. For Hindi, this led to a reduction from 1.3 million audios to 1 million audios. From the discarded audios, audios with average CTC alignment scores greater than 0.9 were added back to the dataset, leading to an addition of 30k more audios for Hindi. Table \ref{tabhour} depicts the number of hours of data left for every language after filtering.

\begin{table}[h]
\small{
\centering
\begin{tabular}{|l|l|}
\hline
\textbf{Language} & \textbf{Hours} \\ \hline
Hindi    &  2913                      \\ \hline
Marathi  & 1962                   \\ \hline
Tamil    & 1190                   \\ \hline
Telugu   & 932                    \\ \hline
Bengali  & 1224                   \\ \hline
\end{tabular}
\caption{Training dataset sizes in no. of hours after filtering}
\label{tabhour}
}
\end{table}
\begin{table*}[]
\small
\centering
\caption{WERs after transcribing generated speech through IndicConformer, where a sentence is 50\% masked}
\label{tab1}
\begin{tabular}{|l|l|c|c|c|} 
\hline
\textbf{Language} & \textbf{Dataset} & \textbf{Ground Truth durations} & \textbf{Infill style durations} & \textbf{Pflow style durations}  \\ 
\hline
\textbf{Tamil}    & IndicSuperb      & 0.2794                          & 0.2806                          & 0.2622                          \\ 
\hline
                  & Mucs             & 0.3438                          & 0.3431                          & 0.3269                          \\ 
\hline
                  & \textbf{overall} & \textbf{0.31160}                & \textbf{0.31185}                & \textbf{0.29455}                \\ 
\hline
\textbf{Telugu}   & Fluers           & 0.4832                          & 0.47                            & 0.4938                          \\ 
\hline
                  & IndicSuperb      & 0.3661                          & 0.4057                          & 0.431                           \\ 
\hline
                  & Mucs             & 0.3842                          & 0.3655                          & 0.3894                          \\ 
\hline
                  & \textbf{overall} & \textbf{0.41117}                & \textbf{0.41373}                & \textbf{0.43807}                \\ 
\hline
\textbf{Bengali}  & Fluers           & 0.4136                          & 0.4391                          & 0.4103                          \\ 
\hline
                  & IndicSuperb      & 0.1948                          & 0.2133                          & 0.1936                          \\ 
\hline
                  & \textbf{overall} & \textbf{0.3042}                 & \textbf{0.3262}                 & \textbf{0.30195}                \\ 
\hline
\textbf{Hindi}    & Fluers           & 0.1797                          & 0.1915                          & 0.2051                          \\ 
\hline
                  & IndicSuperb      & 0.127                           & 0.1542                          & 0.1574                          \\ 
\hline
                  & Mucs             & 0.2022                          & 0.2334                          & 0.2368                          \\ 
\hline
                  & \textbf{overall} & \textbf{0.16963}                & \textbf{0.19303}                & \textbf{0.19977}                \\ 
\hline
\textbf{Marathi}  & Fluers           & 0.3962                          & 0.348                           & 0.3526                          \\ 
\hline
                  & IndicSuperb      & 0.1896                          & 0.1862                          & 0.39                            \\ 
\hline
                  & Mucs             & 0.1214                          & 0.1381                          & 0.351                           \\ 
\hline
                  & \textbf{overall} & \textbf{0.23573}                & \textbf{0.22410}                & \textbf{0.36453}                \\
\hline
\end{tabular}
\end{table*}

\subsection{Model}
Following Voicebox we employed a 103 M parameter model audio model.  Transformer model based on the architecture proposed by \cite{vaswani2023attentionneed}, consisting of 12 layers. Each layer utilizes multi-head self-attention with 16 heads and 512 hidden dim of feed-forward network. Positional information is incorporated using Rotary Positional Embedding (RoPE) \cite{su2023roformerenhancedtransformerrotary} applied within the self-attention mechanism. We used RMSNorm \cite{zhang2019rootmeansquarelayer} for layer normalization. Additionally, the model incorporates UNet like skip connections \cite{ronneberger2015unetconvolutionalnetworksbiomedical} where outputs from the first half of the layers are concatenated and linearly combined with the inputs of the corresponding layers in the second half. 

The Voicebox style duration model uses the same model as audio model with 16 heads, 512 embedding/FFN dimensions, with 12 layers for all our models. 51M parameters in total.  

For the Speaker prompted duration predictor
we employ a Speech-prompted Text Encoder inspired by \cite{kim2023pflow}, and a convolution-based Duration Predictor.
The Text Encoder processes input speech prompt and text embeddings. These are first linearly projected to a shared feature space and concatenated. A prenetwork of three convolutional layers processes this combined sequence. The prenetwork output is split into corresponding prompt and text segments. To differentiate modalities and provide positional context, we add positional encodings to each segment, defined as the sum of standard absolute positional encodings and a unique learnable embedding for prompt or text. These processed representations are then fed into a 12-layer Transformer (8 heads, 512 hidden dimension). The Transformer's attention mechanism is configured to enable text tokens to attend to the speech prompt tokens. The Text Encoder's output provides a speaker-conditional hidden representation.

The Duration Predictor is a shallow convolutional network. It takes the speaker-conditional hidden representation produced by the Text Encoder (prior to its final linear projection) to determine token durations. The total parameters of this model are 84M
All models are trained in FP32.

\subsection{Training}
The audio models are trained for updates 750K/1M steps depending on the language with an effective batch
size of 256k frames. For training efficiency, audio length is capped at 1,000 frames and chunked
randomly if the length exceeds this threshold.
Duration models are trained for 200K updates with an
effective batch size of 200K frames. The AdamW optimizer is used with a peak
learning rate of 2e-4, linearly warmed up for 5K steps and linearly decays over the rest of training.
To encourage robustness of audio model, a probabilistic masking strategy is applied to the input mel representations. With a 50\% probability, a random percentage r of the sequence tokens is masked, where r is sampled from a uniform distribution U[30,100]. With the remaining 50\% probability, either the entire sequence is masked (with 90\% probability, resulting in 45\% overall) or no masking is applied (with 10\% probability, resulting in 5\% overall).   

\subsection{Evaluation Metrics}
\label{sec:evaluation_metric}

\subsubsection{Speaker similarity:} 
To ensure that the generated audio was in the prompt audio speaker's voice, we used various metrics to analyze the the speaker similarity of the prompt and the generated audio.
\begin{enumerate}
    \item \textbf{Sim-o:}  The ECAPA-TDNN \cite{Desplanques-2020} model, a speaker verification model trained on the Voxceleb2 dataset \cite{Chung-2018}, was used to obtain embeddings for the prompt and generated audio. Then, cosine similarity was calculated between these embeddings to obtain a similarity score. Higher sim-o scores indicate higher speaker similarity of two audio pairs.
    \item \textbf{Similarity Mean Opinion Score (SMOS):} 30 human evaluators were employed per language to score the similarity between the prompt and generated audio on a scale of 1 to 5, with 1 being completely dissimilar and 5 being exactly the same. We randomly picked 40 audios for scoring and each audio was annotated by 10 people. Appendix A contains the specific instructions and scoring guidelines given to the humans for SMOS.
\end{enumerate}
\subsubsection{Intelligibility}
To ensure the quality of generated audios, we used several metrics to evaluate their intelligibility.
\begin{enumerate}
    \item \textbf{Word Error Rate(WER):} The IndicConformer \cite{ai4bharat2024indicconformer} model was used to transcribe the generated audios and WER was calculated between these transcriptions and the ground truth transcriptions. A higher WER indicates a poor quality generation.
    \item \textbf{Quality Mean Opinion Score (QMOS):} 30 human evaluators were employed to score the quality of the generated audio on a scale of 1 to 5, with 1 being commpletely unintelligible and 5 being perfectly coherent. We randomly picked 40 audios for scoring and each audio was annotated by 10 people. Appendix A contains the specific instructions and scoring guidelines given to the humans for QMOS.
\end{enumerate}

\section{Results}
\label{sec:results}

The input to our Voicebox audio model was a speaker-specific speech prompt and ground-truth text for text generation. For the text, audio durations were obtained from ground truth durations, P-Flow style speaker-prompted durations and Voicebox-style infil durations. These durations were then used to generate speaker specific speech using the audio model. To measure intelligibility, WERs were calculated for the generated speech. To measure speaker similarity of the generated and original speech, Sim-o scores between the two audio parts were calculated.

\subsection{Continuous Sentence Completion}

We used each instance of the test set as a separate testing sample. Each sentence was split into two halves, with the latter half masked. The first half served as the speaker-specific prompt, while the ground-truth text of the masked second half was provided as the text input. Table \ref{tab1} depicts the dataset-wise WERs for the generated speech from every duration prediction technique. To mitigate the bias introduced by errors from the ASR model used for transcription, we also compare WERs of audios using ground-truth forced aligned durations. The P-flow duration predictor offers gains over Infil-style durations for Tamil and Bengali; however, its performance degrades significantly for Marathi. Table \ref{tab:sim_scores} depicts the Sim-o scores between the original and generated audios for every duration prediction technique. We can observe that the P-flow duration predictor gives improvements over the infil-style durations for Tamil, Telugu and Bengali. 

\begin{table}[h!]
\small{
\centering
\begin{tabular}{l|c|c|c}
\hline
\textbf{Language} & \textbf{GT Sim-o} & \textbf{Infil Sim-o} & \textbf{PFlow Sim-o} \\
\hline
Marathi & 0.6539 & 0.6481 & 0.6393 \\
Tamil   & 0.6902 & 0.6833 & 0.6925 \\
Telugu  & 0.6178 & 0.6292 & 0.633  \\
Bengali & 0.6682 & 0.6573 & 0.6706 \\
Hindi   & 0.6512 & 0.6466 & 0.6344 \\
\hline
\end{tabular}
\caption{Similarity scores across languages for continuous sentence completion of GT, Infil, and PFlow systems}
\vspace{-20pt}
\label{tab:sim_scores}
}
\end{table}

\begin{comment}
    
\subsection{Cross Sentence zero-shot TTS}

An audio was used as the speaker specific prompt, and the text of a completely different, unrelated sentence by a different speaker was used as the text input. Table \ref{tab2} depicts the dataset-wise WERs for the generated speech from every duration prediction technique.  We can observe that the P-flow duration predictor gives improvements over the infil-style durations for the Tamil language. Table \ref{tab:sim_scores2} depicts the Sim-o scores between the original and generated audios for every duration prediction technique. We can observe that the P-flow duration predictor gives improvements over the infil-style durations for Marathi, Tamil, Telugu and Bengali. 

\begin{table}[h!]
\small{
\centering
\begin{tabular}{l|c|c|c}
\hline
\textbf{Language} & \textbf{GT Sim-o} & \textbf{Infil Sim-o} & \textbf{PFlow Sim-o} \\
\hline
Marathi & 0.6897 & 0.7111 & 0.664 \\
Tamil   & 0.7886 & 0.7498 & 0.7925 \\
Telugu  & 0.6938 & 0.6766 & 0.6957 \\
Bengali & 0.7505 & 0.714  & 0.7536 \\
Hindi   & 0.7514 & 0.723  & 0.7493 \\
\hline
\end{tabular}
\caption{Similarity scores across languages for cross sentence zero-shot TTS of GT, Infil, and PFlow systems}
\label{tab:sim_scores2}
}
\end{table}

\end{comment}

\subsection{Human Evaluations}
Human evaluations were used to get SMOS and QMOS scores. Detailed human annotation instructions are in Appendic A and B. Table \ref{tab:my-table} contains SMOS and QMOS scores which show that the speaker-prompted duration predictor performs better than the infilling duration predictor for Hindi and Tamil.
\begin{table}[h]
\small{
\setlength{\tabcolsep}{3.5pt}
\begin{tabular}{|l|l|l|r|r|r|}
\hline
\textbf{Lang.}        &  \textbf{Score}             & \textbf{}                & \multicolumn{1}{l|}{\textbf{GT}} & \multicolumn{1}{l|}{\textbf{Infill}} & \multicolumn{1}{l|}{\textbf{Pflow}} \\ \hline
\textbf{Hindi}   & QMOS          & Naturality               & 4.3449                           & 3.9742                               & 4.3349                     \\ \hline
                 &               & Intelligibility          & 4.4646                           & 4.2397                               & 4.1313                     \\ \hline
                 & SMOS          & Similarity               & 3.9867                           & 3.57                                 & 4.1356                     \\ \hline
                 &               & \textbf{Overall}         & \textbf{4.2654}                  & \textbf{3.9280}                 & \textbf{4.2006}            \\ \hline
\textbf{Tamil}   & QMOS & Naturality      & 4.4277                  & 4.4792                      & 4.4856                     \\ \hline
\textbf{}        &               & Intelligibility          & 4.4378                           & 4.4682                               & 4.6063                     \\ \hline
                 & SMOS          & Similarity               & 4.5343                           & 4.4121                               & 4.5619                     \\ \hline
                 &               & \textbf{Overall}         & \textbf{4.4666}                  & \textbf{4.4532}                 & \textbf{4.5513}       \\ \hline
\textbf{Bengali} & QMOS          & Naturality               & 4.1688                           & 4.2678                               & 4.1125                     \\ \hline
\textbf{}        & \textbf{}     & Intelligibility & 4.4538                  & 4.4748                      & 4.3788                     \\ \hline
\textbf{}        & SMOS          & Similarity               & 3.8065                           & 3.9685                               & 3.9176                     \\ \hline
                 &               & \textbf{Overall}         & \textbf{4.1430}             & \textbf{4.2370}                 & \textbf{4.1363}            \\ \hline
\end{tabular}
\caption{QMOS and SMOS scores for Hindi, Tamil and Telugu outputs.}
\label{tab:my-table}
}
\end{table}

\section{Conclusion}

We have performed an analysis of different techniques to obtain speaker-specific durations during zero-shot speaker-specific TTS. We analyzed a Voicebox-style infillling duration predictor and a P-Flow-style speaker prompted duration predictor. It was observed that the speaker prompted duration predictor offered considerable benefits over the infilling duration predictor for Tamil, in both intelligibility and speaker similarity. Additionally as a general trend, the speaker-prompted duration predictor led to better speaker similarity, while the infilling duration predictor led to better intelligibility. This trade-off gives us insights into the importance of the duration prediction process and how it can affect not only the intelligibility, but also the speaker similarity of generated audios. We can leverage this knowledge to choose a suitable duration predictor based on our use-case, and also to develop a new duration predictor that adapts properties of both speaker-prompted and infilling-style duration predictors to improve both intelligibility and speaker-similarity.

\bibliography{mybib}

\section*{Appendix A: QMOS guidelines}

\subsection{Naturalness Evaluation}

\textbf{Definition:} Naturalness refers to how lifelike and fluid the synthesized speech sounds.

\begin{table}[h]
\small{
\centering
\begin{tabular}{|c|p{5cm}|}
\hline
\textbf{Score} & \textbf{Description} \\ \hline
5 & Completely natural, indistinguishable from a human speaker. \\ \hline
4 & Mostly natural, but with slight unnatural elements. \\ \hline
3 & Moderately natural, with noticeable synthetic artifacts or monotony. \\ \hline
2 & Mostly unnatural, robotic or artificial-sounding. \\ \hline
1 & Completely unnatural, heavily robotic, or difficult to listen to. \\ \hline
\end{tabular}
\caption{Description of naturalness scores.}
\label{tab:naturalness}
}
\end{table}

\subsection{Intelligibility Evaluation}

\textbf{Definition:} Intelligibility measures how easily the speech can be understood, regardless of how natural it sounds. It focuses on clarity and accuracy of pronunciation.

\begin{table}[h]
\small{
\centering
\begin{tabular}{|c|p{5cm}|}
\hline
\textbf{Score} & \textbf{Description} \\ \hline
5 & Perfectly clear; every word is easily understood. \\ \hline
4 & Mostly clear, but with minor pronunciation errors or distortions. \\ \hline
3 & Somewhat clear; requires some effort to understand certain words. \\ \hline
2 & Mostly unclear; many words are difficult to recognize. \\ \hline
1 & Completely unintelligible; nearly impossible to understand. \\ \hline
\end{tabular}
\caption{Description of intelligibility scores.}
}
\end{table}

\subsection{Guidelines}

\begin{enumerate}
    \item Listen to each audio sample carefully. Replay if necessary.
    \item Rate the sample separately for Naturalness, Intelligibility, and Speaker Similarity on a scale from 1 to 5.
    \item Avoid bias by focusing on the specific criteria, not personal preference.
    \item Explain the score clearly. Justify the score by describing key factors such as errors, inconsistencies, or deviations from the expected standard.
\end{enumerate}

\subsection{Important guidelines}

\begin{itemize}
    \item Use headphones for better sound quality.
    \item Ensure a quiet environment to avoid distractions.
    \item Rate objectively without comparing different speech styles or accents.
    \item Do not assume meaning—rate based on what you actually hear.
\end{itemize}

\textbf{Thank you for your contribution!}

\section*{Appendix B: SMOS guidelines}

\subsection{Speaker Similarity Evaluation}

\textbf{Definition:} Speaker similarity refers to how much the synthesized voice resembles the reference speaker’s voice in terms of timbre, pitch, and prosody. Ignore other factors like intelligibility and only focus on speaker similarity.

\begin{table}[h]
\small{
\centering
\begin{tabular}{|c|p{5cm}|}
\hline
\textbf{Score} & \textbf{Description} \\ \hline
5 & Indistinguishable from the reference speaker. \\ \hline
4 & Very similar, but with minor differences. \\ \hline
3 & Moderately similar, but noticeable variations. \\ \hline
2 & Weak similarity, with clear differences in voice identity. \\ \hline
1 & Completely different from the reference speaker. \\ \hline
\end{tabular}
}
\end{table}

\subsection{Guidelines}

\begin{enumerate}
    \item Listen to each audio sample carefully. Replay if necessary.
    \item Rate the sample separately for Naturalness, Intelligibility, and Speaker Similarity on a scale from 1 to 5.
    \item Avoid bias by focusing on the specific criteria, not personal preference.
    \item Explain the score clearly. Justify the score by describing key factors such as errors, inconsistencies, or deviations from the expected standard.
\end{enumerate}

\subsection{Important guidelines}

\begin{itemize}
    \item Use headphones for better sound quality.
    \item Ensure a quiet environment to avoid distractions.
    \item Rate objectively without comparing different speech styles or accents.
    \item Do not assume meaning—rate based on what you actually hear.
\end{itemize}

\section*{Appendix C: Test set creation}
We use the Vistaar test set for all our benchmarking. From this set, we select audio clips ranging from 3 to 10 seconds in duration. Test sets yielding fewer than 100 instances after this filtering are excluded. Table~\ref{tab:lang-dataset-sizes} lists the datasets and the number of samples used for benchmarking after filtering.
\begin{comment}
\begin{table}[t]
\centering
\scriptsize
\caption{Number of samples per dataset for each language.}
\label{tab:lang-dataset-sizes}
\begin{tabular}{|l|l|r|}
\hline
\textbf{Language} & \textbf{Dataset}   & \textbf{Samples} \\
\hline
\multirow{4}{*}{Tamil}   & indicvoices      & 3     \\
                         & IndicSuperb      & 2598  \\
                         & Mucs             & 1137  \\
                         & \textbf{Total}   & \textbf{3738} \\
\hline
\multirow{4}{*}{Telugu}  & Fluers           & 374   \\
                         & IndicSuperb      & 2892  \\
                         & Mucs             & 1166  \\
                         & \textbf{Total}   & \textbf{4432} \\
\hline
\multirow{4}{*}{Bengali} & indicvoices      & 13    \\
                         & Fluers           & 787   \\
                         & IndicSuperb      & 2898  \\
                         & \textbf{Total}   & \textbf{3698} \\
\hline
\multirow{5}{*}{Hindi}   & indicvoices      & 17    \\
                         & Fluers           & 360   \\
                         & IndicSuperb      & 1876  \\
                         & Mucs             & 917   \\
                         & \textbf{Total}   & \textbf{3170} \\
\hline
\multirow{5}{*}{Marathi} & indicvoices      & 6     \\
                         & Fluers           & 825   \\
                         & IndicSuperb      & 1006  \\
                         & Mucs             & 316   \\
                         & \textbf{Total}   & \textbf{2153} \\
\hline
\end{tabular}
\end{table}
\end{comment}

\begin{table}[t]
\centering
\caption{Number of samples per dataset for each language.}
\label{tab:lang-dataset-sizes}
\begin{tabular}{|l|l|r|}
\hline
\textbf{Language} & \textbf{Dataset}   & \textbf{Samples} \\
\hline
\multirow{4}{*}{Tamil}  
                         & IndicSuperb      & 2598  \\
                         & Mucs             & 1137  \\
                         & \textbf{Total}   & \textbf{3738} \\
\hline
\multirow{4}{*}{Telugu}  & Fluers           & 374   \\
                         & IndicSuperb      & 2892  \\
                         & Mucs             & 1166  \\
                         & \textbf{Total}   & \textbf{4432} \\
\hline
\multirow{4}{*}{Bengali} 
                         & Fluers           & 787   \\
                         & IndicSuperb      & 2898  \\
                         & \textbf{Total}   & \textbf{3698} \\
\hline
\multirow{5}{*}{Hindi}  
                         & Fluers           & 360   \\
                         & IndicSuperb      & 1876  \\
                         & Mucs             & 917   \\
                         & \textbf{Total}   & \textbf{3170} \\
\hline
\multirow{5}{*}{Marathi} 
                         & Fluers           & 825   \\
                         & IndicSuperb      & 1006  \\
                         & Mucs             & 316   \\
                         & \textbf{Total}   & \textbf{2153} \\
\hline
\end{tabular}
\end{table}
\textbf{Thank you for your contribution!}

% \section{Appendix}

% \subsection{Comparison with F5TTS}
% \textcolor{red}{ISHA complete} \\
% F5TTS takes a long time to converge. On 100 hours A100 40 GB  8GPUs it took a week. 
% While voicebox A100 80GB takes 3 days on 3k hours of data. 
% 200 hours 8GPUS H100 80 GB - On train

% Advaids: dim: 768, depth=18, 12 heads 100 hours 1.2 m (IndicTTS and IndicVoices)
% Ours: dim: 1024, depth=22, 16 heads, 3k hours 3-4 days
% 200 hours 4 days - Listen some words here and there (IndicVoices and Indic Supberb)

% 24 Khz - paper
% 16 Khz - our

% \textcolor{red}{Will contain reasoning why we didnt go with F5TTS, RTF and training convergence time comparison as well as wer, simo}

% \subsection{Human Evaluation}
% \textcolor{red}{Will contain specifics of human scoring guidelines for both QMOS and SMOS}

% \subsection{Dataset specific analysis}
% \textcolor{red}{Will contain table of with and without filtering and separate datasets, only for hindi is enough}

% \subsection{WER evaluation}
% \textcolor{red}{Will contain why whisper was chosen over wav2vec- small comparison of GT wers - can be ditched}

\end{document}